\documentclass[fleqn,usenatbib]{mnras}

\usepackage{txfonts}

\usepackage[T1]{fontenc}

\DeclareRobustCommand{\VAN}[3]{#2}
\let\VANthebibliography\thebibliography
\def\thebibliography{\DeclareRobustCommand{\VAN}[3]{##3}\VANthebibliography}

\usepackage{gensymb}
\usepackage{graphicx}	
\usepackage{amsmath}	
\usepackage{amssymb}	

\title[PyNAPLE: Lunar Surface Impact Crater Detection]{PyNAPLE: Lunar Surface Impact Crater Detection}

\author[D. Sheward et al.]{
D. Sheward,$^{1}$\thanks{E-mail: djs22@aber.ac.uk (DS)}
C. Avdellidou,$^{2,3}$
A. Cook,$^{1}$
E. Sefton-Nash,$^{3}$
M. Delbo,$^{2}$
B. Cantarella,$^{4}$
and L. Zanatta$^{4}$\\
$^{1}$Department of Physics, Aberystwyth University, Ceredigion, SY23 3BZ, UK \\
$^{2}$Universit\'e C\^ote d'Azur, Laboratoire Lagrange, Observatoire de la C\^ote d'Azur, CS 34229-F 06304 Nice Cedex 4, France\\
$^{3}$ESTEC, European Space Agency, Keplerlaan 1, 2201 AZ Noordwijk, The Netherlands \\
$^{4}$Sezione di Ricerca Luna dell'Unione Astrofili Italiani\\}

\date{Accepted XXX. Received YYY; in original form ZZZ}

\pubyear{2021}

\begin{document}
\label{firstpage}
\pagerange{\pageref{firstpage}--\pageref{lastpage}}
\maketitle

\begin{abstract}

In the last 20 years, over 600 impact flashes have been documented on the lunar surface. This wealth of data presents a unique opportunity to study the meteoroid flux of the Earth-Moon environment, and in recent years the physical properties of the impactors. However, other than through serendipitous events, there has not been yet a systematic search and discovery of the craters associated to these events. Such a meteoroid-crater link would allow us to get insight into the crater formation via these live observations of collisions. Here we present the PyNAPLE (Python NAC Automated Pair Lunar Evaluator) software pipeline for locating newly formed craters using the location and epoch of an observed impact flash. We present the first results from PyNAPLE, having been implemented on the 2017-09-27 impact flash. A rudimentary analysis on the impact flash and linked impact crater is also performed, finding that the crater's ejecta pattern indicates an impact angle between 10-30\degree, and although the rim-to-rim diameter of the crater is not resolvable in current LRO NAC images, using crater scaling laws we predict this diameter to be 24.1-55.3~m, and using ejecta scaling predict a diameter of 27.3-37.7~m. We discuss how PyNAPLE will enable large scale analyses of sub-kilometer scale cratering rates and refinement of both scaling laws, and the luminous efficiency. 

\end{abstract}




\section{Introduction}

Impact craters have been observed on several types of solar system bodies, such as planets, natural satellites, asteroids, comets, and transneptunian objects. Moreover, the larger atmosphereless bodies such as Mercury, the Moon, and Ganymede have impact craters covering the majority of the surface. By studying these impact craters we aim to understand how the most violent process of impacts shape the planetary surfaces, reveal the chronology of surfaces \citep{neukum2001,hartmann2005}, as well as the mixing of materials on them \citep{dellagiustina2021,tatsumi2021, avdellidou2018}.

Cratering studies are predominately done in hypervelocity impact laboratories where the impact conditions can be controlled, such as the impact speed, impact angle, and the materials of the target and impactor. Moreover, in the laboratory the impact outcome can also be monitored both during and post-impact such as the amount and speed of the ejecta \citep{housen2011,housen2012}, the size and morphology of the crater \citep{housen1999,housen2003, housen2018,avdellidou2020}, the survival of the projectile \citep{avdellidou2016,avdellidou2017, daly2015, daly2016}, etc. However, laboratory experiments are restricted by the relatively low speeds \citep[<8~km~s$^{-1}$,][]{burchell1999} and the small size of the projectile (a few mm) that can be used as an impactor at such speeds. 

A number of crater scaling laws have also been developed which describe the relationship between the impactor energy and the size of the resulting crater \citep{horedt1984, melosh1989}. These crater scaling laws however are of questionable accuracy; each scaling law proposed is generally accurate only for the energy, mass, and velocity ranges of the experiments from which the laws were derived, becoming increasingly inaccurate as these ranges are deviated from.

The study of the impacts themselves have been the subject of several missions. ISRO's Chandrayaan-1 Moon Impact Probe \citep{goswami2009}, and NASA's Lunar Crater Observation and Sensing Satellite (LCROSS) \citep{lcross} missions were both impacted into the Lunar surface in order to observe for volatiles in the ejecta cloud, and the formed impact craters have been subsequently studied using NASA Lunar Reconnaissance Orbiter (LRO) data \citep{hayne2010,colaprete2010,schultz2010}. Furthermore, the sample return mission Hayabusa2 (JAXA) performed an artificial impact on the surface of the near-Earth asteroid Ryugu \citep{arakawa2020}. Nevertheless such large-scale experiments are rare and expensive for systematic studies. 

Every year the Earth is bombarded by over 12000 tonnes of rock, metal, and ice, from space in the form of meteoroids \citep{drolshagen2017}. While the majority of this material ablates in the atmosphere as a meteor, this material can cause substantial damage through both direct collisions with the ground, and through exploding as an air-burst such as in Chelyabinsk in 2013 \cite[see e.g.][and references therein]{popova2013}, and Tunguska in 1908 \cite[see e.g.][and references therein]{bagnall1988}. It is estimated that the frequency of impacts on Earth by a cm-size object is 1000 times greater than a meter-size one \citep{suggs2014}. However, such small impactors never reach the ground to form a crater, due to them completely ablating in the atmosphere. For this reason, in the last 20 years the attention is shifted towards the atmosphereless Moon, where the most common impact study is the monitoring for transient luminous event caused by impacting meteoroids, known as lunar impact flashes (LIF). During an impact event the kinetic energy ($KE$) of the impactor is partitioned, and a small fraction (<0.5\%) is transformed to luminous energy ($E_{lum}$). 

These LIF are bursts of light, typically lasting a few milliseconds but are detectable through moderately sized telescopes, are therefore available for observing by both professionals and amateur astronomers alike. These LIFs have been observed since 1999 \citep{ortiz1999, ortiz2000}, with unconfirmed reports as early as 1953 \citep{johnson2001}, and in total over 600 LIFs have been reported in the literature or in LIF databases \citep{ortiz2002,ortiz2006,ortiz2015,madiedo2014,madiedo2015,madiedo2018,larbi2015,bonanos2018, zuluaga2020, avdellidou2021}. 

There have been analysis methods developed where several parameters of the impact process itself \citep{suggs2014,madiedo2015b,avdellidou2021} as well as physical properties of the impactors \citep{avdellidou2019,avdellidou2021} can be determined. These methods allowed the construction of the size frequency distribution of the cm-dm meteoroids \citep{suggs2014,avdellidou2021} as well as the distribution of the temperatures \citep{avdellidou2019,avdellidou2021} of the flash events that appear to verify the theoretical estimations \citep{cintala1992,nemtchinov1998}. The identification of the craters caused by the impacts observed as LIFs is of a paramount importance, in order to establish the desired link between the size of the impactor and the size of the crater. In the recent past, the Lunar Reconnaissance Orbiter discovered over 220 fresh craters \citep{speyerer2016}. Only two fresh impact craters, however, have been located following LIF observations; one by the Marshall Space Flight Center on 2013-03-17 \citep{moser2014,robinson2015}, and the other by MIDAS on 2013-09-11 \citep{madiedo2014,crater2}. The detection of the first crater was serendipitous rather than systematic, and the detection of the second crater involved performing targeted LRO observations to aid in their search - a process which is not viable for the general scientific community. 

In this work we present the PyNAPLE (Python Nac Automated Pair Lunar Evaluator) algorithm pipeline, developed to use LIF observations and subsequently identify new craters on the Moon using the images from the LRO. PyNAPLE utilises broadly the same methodology as described by \cite{speyerer2016}, however applying it in a more targeted manner. PyNAPLE has been designed with the intention of being utilised as a publicly available tool for both amateurs and professionals to use, and its primary aim is to produce a substantial data set of impact flash linked impact craters. 

In section 2 we discuss the datasets that PyNAPLE is designed to utilise, which are also used to calibrate and test the algorithm. In section 3 we discuss its operating procedure, section 4 discusses the tests performed on PyNAPLE, and section 5 presents the first result of a confirmed crater. In section 6 we perform a short analysis on the crater and we discuss how PyNAPLE will enable larger scale statistical analyses of impact crater formation at the sub-metre impactor scale, with the intent of utilising this data to refine crater scaling laws, and examine the accuracy of the luminous efficiency values used for lunar impacts.

\section{Datasets}

\subsection{Images from LROC Instrument}

The Lunar Reconnaissance Orbiter Camera (LROC) consists of a Wide Angle Camera (WAC), and two Narrow Angle Cameras (NAC) \citep{robinson2010}. These NACs are panchromatic line-scan cameras with 5064 pixels, which combine to give 0.5~m~px$^{-1}$ resolution at 50~km altitude. As of February 2022, the LROC has obtained over 3 million images, totalling over 1 petabyte of data. The LROC began its imaging of the lunar surface 2009-09-15, and is expected to keep operating at least until 2026, giving a 17-year span over which impact flash linked craters can be detected by PyNAPLE. 

LROC images are released publicly on a monthly basis approximately three months after collection, giving the minimum time between observation to crater detection of 3 months. While the imaging performed by the LRO is not continuous, having readout times between subsequent images, the area covered by the LRO WAC in the month-long image release period generally encompasses the whole Moon. The surface area covered by LRO NAC images in this time is approximately an order of magnitude less than the area the LRO WAC covers, having only partial coverage of the lunar surface in a stochastic pattern. Due to the nature of the LRO mission having targets of interest for which the spacecraft will re-orient to observe, and inconsistent capture and readout times for images, the gap between LRO NAC images can be erratic. The ability to predict whether a certain point on the lunar surface will be imaged in a given month is therefore not possible without knowledge of the LROs target schedule.

PyNAPLE utilises LRO NAC images as its basis for locating changes. While WAC images offer greater surface coverage of the lunar surface, its nominal resolution of 100~m~px$^{-1}$ is not sensitive enough for the majority of impact crater detection; the ejecta blankets caused by the most common observable impact flashes would be less than a pixel in size at this scale. NAC images are instead used, and provide almost total coverage between 75\degree and -75\degree, with the sub-metre resolution being more suitable to resolve finer details of the typically >50~m craters. Using this large dataset of images provides a consistency between images which aids to simplify image processing and prevent any artefacts produced from using dissimilar images collected by differing instruments. 

\subsection{Lunar Impact Flash Observations}

The main function of PyNAPLE is to search for new craters at locations where impacts have been recorded on the Moon via the observation of their impact flashes. 

Several observational campaigns exist for the purpose of observing and recording LIFs. NASAs Meteoroid Environment Office (MEO) runs an observational campaign since 2005 and collates observed events into a publicly available list of events\footnote{https://www.nasa.gov/centers/marshall/news/lunar/lunar\_impacts.html}, however, the actual video frames are not made available \citep{suggs2008}. Over 120 events have been analysed and published by their team \citep{suggs2014}.
NELIOTA is an ESA-funded project running since 2017 and provides publicly available data\footnote{https://neliota.astro.noa.gr} from the LIFs observed \citep{xilouris2018}. MIDAS is a Spanish based observation campaign that releases findings in published works. The MIDAS team developed their own software for LIF detection and analysis which is not publicly available \citep{madiedo2015b}. While these campaigns each have different methods of observing, the general concepts are the same; observing during local night using mostly visible systems to monitor the lunar night-side. Currently, only NELIOTA provides observations detected simultaneously in visible (R-Band) and near-infrared (I-Band) wavelength range.

LIFs are usually reported together with the time of the impact, the peak magnitude and the duration of the flash as well as the selenographic coordinates where it occurred. These publicly available LIF observations are the input in PyNAPLE, which allows searches for the resultant impact craters.

\section{PyNAPLE Algorithm}

\begin{figure*}
	\includegraphics[width=\textwidth]{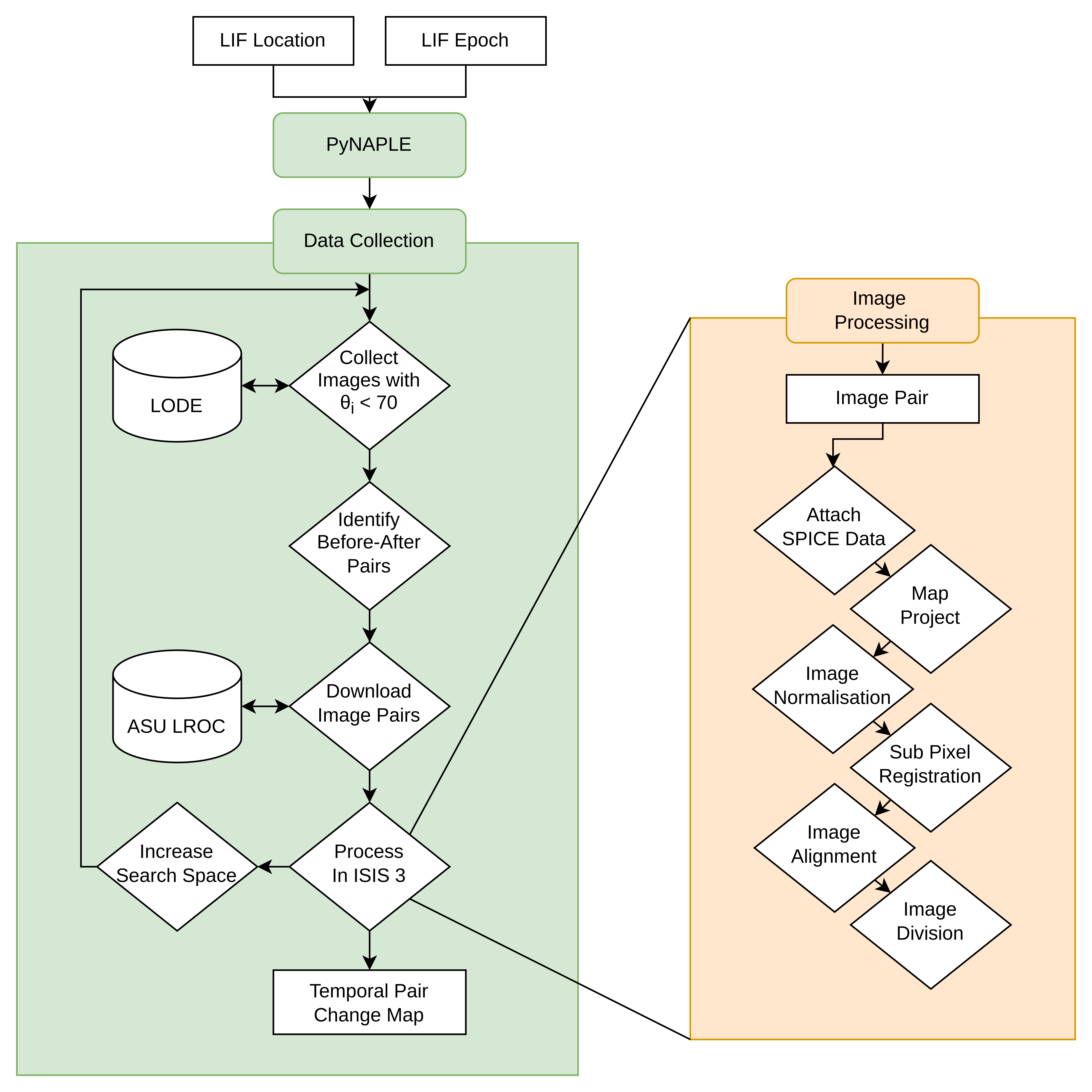}
    \caption{The PyNAPLE algorithm works in two distinct sections, data collection and image processing, and utilises the publicly-available online LROC image repositories; the \textit{Lunar Orbital Data Explorer}, and ASUs \textit{LROC Image Search}, and the image processing capabilities of USGSs ISIS3. Rectangular boxes are data, diamond boxes are processes, and cylindrical boxes are online resources.}
    \label{fig:flow}
\end{figure*}

PyNAPLE was designed with three main focuses; simplicity, full automation, and configurability. 
In order to meet the first two components, PyNAPLE has been built to run with only a latitude, longitude, and epoch of a LIF, and run until all temporal pairs have been processed within the search space without the need for further user input, using the optimised default values. Moreover, the configurability of PyNAPLE is important, and therefore tolerances, matching algorithms, and search space size and increment are all able to be altered from the default through the use of a the configuration file. For the purposes of discussion, the default values are used within this section.

The PyNAPLE algorithm works in two stages, data collection, and data processing. The data collection stage requires an online connection, as PyNAPLE utilises online repositories to collate the image and SPICE data. The data processing step is implemented using the United States Geological Survey (USGS) Integrated Software for Imagers and Spectrometers (ISIS3) to process the images \citep{isis3}. PyNAPLE serves as a \texttt{Python} based wrapper for these two stages, serving to simplify and automate the entire process. The general flow of the PyNAPLE algorithm is shown in Fig.~\ref{fig:flow}.

The data collection stage of PyNAPLE utilises two publicly available online LROC data repositories to generate the image lists, the Lunar Orbital Data Explorer\footnote{ode.rsl.wustl.edu/moon/indexProductSearch.aspx} (LODE) \citep{lode}, and Arizona State University's (ASU) LROC Image Search\footnote{https://wms.lroc.asu.edu/lroc/search}. The decision was made to operate this way for two reasons. Firstly, PyNAPLE is already resource intensive for both storage and processing - by using an online service the amount of local data could be kept to a minimum. Secondly, the nature of the two online repositories meant both were faster for different tasks - LODE is faster for looking up the ephemeris data and parameters of all the images in a given area, and ASU LROC Image Search is quicker for downloading the required images. An added benefit of this decision is that it limits the number of requests to either service, and therefore keeps the server load caused by PyNAPLEs requests distributed and at a minimum. 

By connecting to LODE, PyNAPLE generates two lists of images of the location, before and after the impact.
These image lists are then checked, and any images with incidence angle, $\theta_i$ >~70\degree with respect to the normal are discarded, in order to minimise the amount of shadows present in the image. These two lists are then checked for temporal pairs that can be formed. 
Temporal pairs are a pair of "before" and "after" images, which can be aligned and divided in order to attenuate areas where no change has occurred, as the pixel digital number, DN$\rightarrow$0, and accentuate details which have changed between images, as DN$\rightarrow$255. 


In order to successfully identify physical changes of the lunar surface, distinguishing them from changes in illumination conditions, images selected for use in temporal pairs are subject to constraints. The change in incidence angle, $\Delta\theta_i$, should be <~15\degree. This is done to minimise the difference in illumination of the images; when $\theta_{iA}$~=~$\theta_{iB}$ shadows within the image will be the similar and cancel out during division. As $\Delta\theta_i$ increases, shadows in the images will start to differ, and therefore the number of false detections will increase. Similarly, the sub-solar longitude with respect to the target longitude - the relative sub-solar longitude, $\lambda_s$, of the two images must be either both positive or both negative, such that $\lambda_{sa} \lambda_{sb}~>~0$. Alternatively $\Delta\lambda$~<~15\degree is accepted in lieu; either one of these thresholds can be met to ensure the direction and size of any shadows present are similar between images.

\begin{figure}
	\includegraphics[width=\columnwidth]{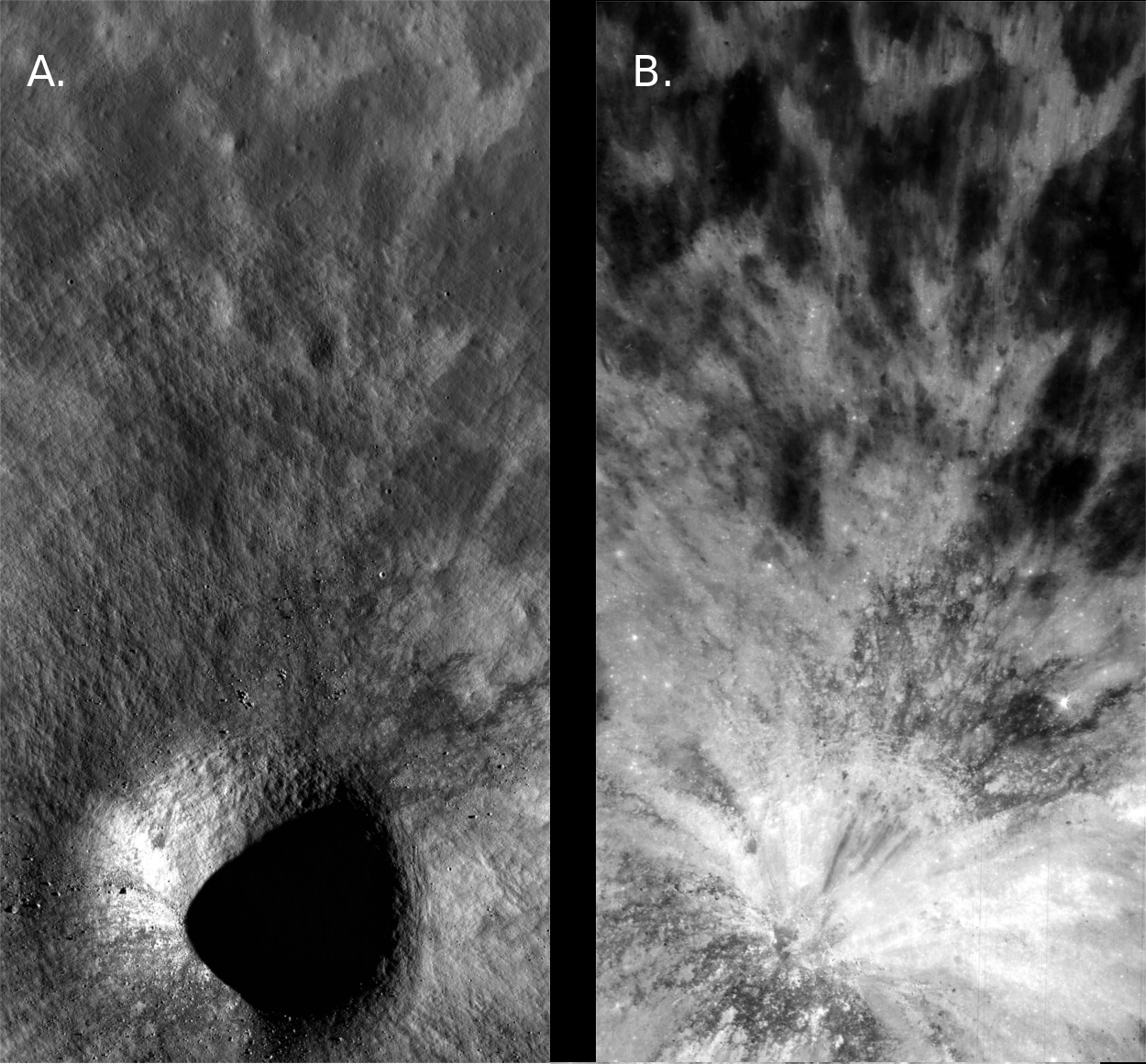}
    \caption{Two LROC images, \textbf{(A)} M146255155RE, and \textbf{(B)} M187520776LE, taken at incidence angles $\theta_i$=68.0\degree, and $\theta_i$=8.3\degree, respectively. While both images are of the same location, the differing illumination causes the images to be visually dissimilar, preventing the formation of a temporal pair. Images cropped from NAC images, NASA/GSFC/Arizona State University.}
    \label{fig:ill_angle}
\end{figure}

These tolerances are important, as forming temporal pairs outside them causes the images to be too observationally different for the "cancellation" effect of the image division to occur. This visual dissimilarity can be seen in Fig.~\ref{fig:ill_angle}. Although \cite{speyerer2016} in their survey implemented the stricter values of $\theta_i$~<~50\degree, and $\Delta\theta_i$~<~3\degree in their search for surface changes, their work took a broader approach to change detection and was more focused on finding any changes which had occurred within the any temporal pairs at any location. Conversely, our work focuses on finding a specific crater within a specific locale, and we thusly define looser tolerances in order to increase the number of potential temporal pairs, giving a higher probability of detection.

Once the temporal pairs are selected in compliance with the constraints described have been identified, the experimental data records (EDRs) for each image are downloaded from Arizona State University's LROC Image Search. 

The image pairs are then processed through several USGS ISIS3 programs. For each pair, both images are first initialised using SPICE data with \texttt{spiceinit} and \texttt{spicefit}, and then calibrated using \texttt{lronaccal} and \texttt{lronacecho}. This ensures that both the images have accurate location and instrument pointing data, and processes the images to remove the average bias, dark current, and remove the channel echo effect discovered in LRO NAC images in 2011 \citep{humm2016}.

The images are then map projected into the same user-specified projection using \texttt{cam2map}, and \texttt{tonematch} is run to normalise the images. A control network is then generated for the image pair, first using \texttt{autoregtemplate} to define the matching algorithm and chip sizes, and then registering the images using \texttt{coreg}. This control network is a list of points on the surface which have been correctly matched between the two images, and serves as the basis for warping the image to remove any inconsistencies due to the camera instrument being in differing locations between images. 

Using \texttt{warp}, the "after" image is warped onto the "before" image, ideally creating an "after" image with the same observational geometry as the "before" image with sub-pixel accuracy. As this accuracy is not ensured at this stage, the registration and warp process is then repeated with smaller chip size and more strict matching algorithm tolerance, in order to bring each pixel in the "after" image into alignment with the "before" image pixel which corresponds to the same part of the lunar surface. Any pixels which do not have a corresponding pixel at this stage are removed, in order to keep file sizes and processing time to a minimum.
 
Once sub-pixel accuracy has been ensured, division can occur using \texttt{ratio} and the "before" image as the denominator, producing an image which highlights any changes that have occurred between the images.
This process is then repeated for each temporal pair in the list. Upon completing the final pair, the search space in which image pairs are evaluated is increased by 0.25\degree in both the latitude and longitude direction. If the search space is not yet at the user-defined maximum, the process will repeat using the larger search space.

\subsection{Dependencies and Requirements}

PyNAPLE has several dependencies on third-party software. Due to the image processing taking place in ISIS3, the base ISIS3 software is required with LRO module and SPICE kernels, which in total occupies approximately 260~Gb. Several \texttt{Python} packages are also required, most of which are ubiquitous, however a list of any missing packages is generated upon first running PyNAPLE.

The size of the NAC images also factors into the storage requirements, each image once having SPICE data attached occupies between 1-2~Gb of space. Due to the large size of the images, and the operations being performed on them, PyNAPLE also requires a fair amount of computing resources. As each image is processed on a single core, the processor speed of the machine greatly affects the speed at which temporal pairs can be processed. 

Due to the resource intensive and pipeline critical step of registering the image pairs with sub-pixel accuracy 8~Gb of RAM is a necessity to handle image registration. While this could be reduced by using a control network containing less control points, this would lower the quality of the image alignment, and is not a favourable trade off. 

\section{Testing}

Before applying PyNAPLE to locate newly formed craters, we first tested the algorithm using an impact crater discovered by the LRO team, following the LIF report by the MEO on 2013-03-17 \citep{suggs2014}.

The first test performed was to input the coordinates and epoch of the 2013-03-17 impact flash, to confirm that the crater could be located under best-case-scenario parameters. In this test, the crater was located in the first temporal pair processed, showing that the algorithm was effective. 
The second test was again performed using the 2013-03-17 impact, however the coordinates were changed so that the crater was 1\degree outside the initial search area in both latitude and longitude. This tested the PyNAPLE algorithms effectiveness when the crater is further away from the supplied coordinates, as is typically expected when using coordinates estimates from an impact flash. 
The third test performed was to provide the exact coordinates of the impact crater to PyNAPLE, however changing the epoch to be later than the crater formation time frame. As there were now images from before the event which contained the crater, the newly formed crater would only appear in some of the temporal pairs, and therefore not be possible to be the linked crater. 

Once PyNAPLE sufficiently passed these tests, the algorithm used publicly available LIF data for which the produced craters have not yet been identified. 

\section{Results}

\subsection{2017-09-27 Impact Flash}
\begin{figure}
	\includegraphics[width=\columnwidth]{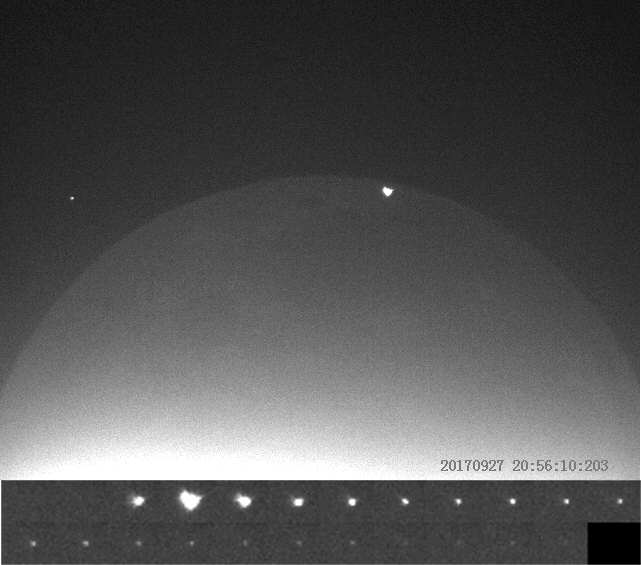}
    \caption{\textbf{Top:} The full frame of the 2017-09-27 impact flash at its peak. \textbf{Bottom:} The cropped sequence of the impact flash from Camera 1.}
    \label{fig:flash_frame}
\end{figure}

At 18:56:12~UT on 2017-09-27, a meteoroid impacted the lunar surface releasing a flash of light observed by La Sezione di Ricerca Luna dell'Unione Astrofili Italiani (SdR UAI), from Melazzo, AL, Italy. The flash was observed by two separate telescope-camera systems, and is shown in Fig.~\ref{fig:flash_frame}. The first system consisted of a 20~cm Newtonian telescope (f/5 reduced to f/2.9) equipped with an ASI120MM CMOS camera with resolution 640$\times$480 pixels, while the second system consisted of a 10~cm Newtonian telescope (f/4) with an ASI120MM running at resolution 512$\times$384 pixels. Both cameras operated with fast acquisition at 25 and 30~fps respectively, and 2$\times$2 binning. No photometric filters were attached to the telescope-camera system, and therefore provided a sensitivity window dictated by the spectral response of the cameras, from approximately 400~nm to 700~nm.

The flash was resolvable above the background noise in 30 frames in the first system, and 24 frames in the second system, giving a minimum duration of approximately 1~s. After applying a Gaussian fit correction to the saturated flash, photometry can be applied to the reconstructed point spread function to obtain the apparent magnitude of the flash. 

While there were no observations of an ideal reference star, Gaia DR2 4094972869712553088, was observable within the flash frames. As this is a variable star with only a small number of scientific observations the exact magnitude of the star at the time of observation is unknown, and therefore the mean value of magnitude~7, as observed by Gaia Data Release 2 \citep{gaiadr2}, was used for photometry. 

\begin{figure}
	\includegraphics[width=\columnwidth]{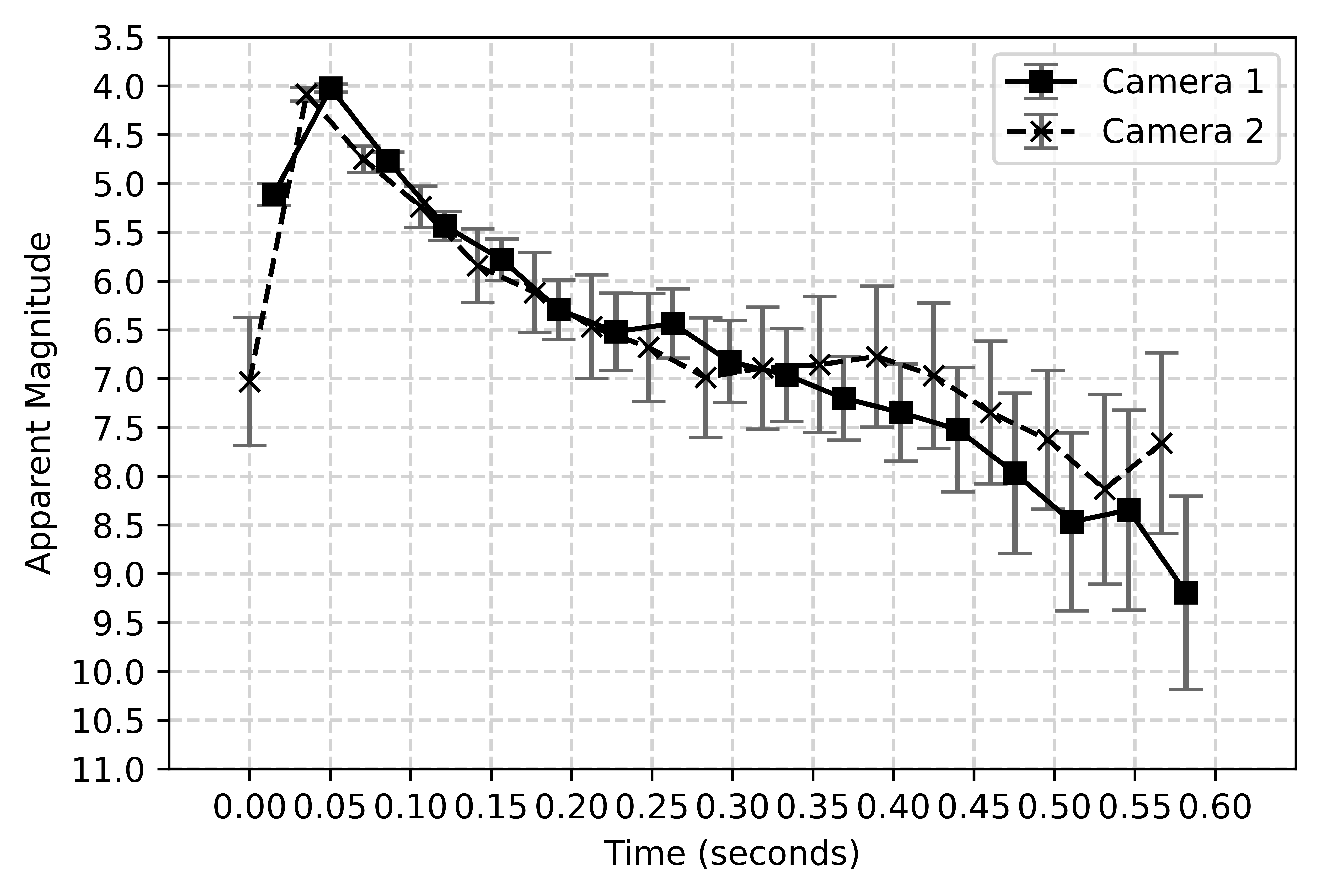}
	\caption{The calculated apparent magnitudes from the first 17 frames of the 2017-09-27 impact flash.}
	\label{fig:mag}
\end{figure}

The evolution of the first 17 frames of the flash from each camera are shown in Fig.~\ref{fig:mag}. Only 17 frames are shown as after this point the signal from the flash is dominated by the noise level, and the calculated values for magnitude are unreliable. Photometry on the Gaussian corrected flash gives the apparent magnitude of the peak as approximately +4.0. Multiple aperture photometry was performed on the Gaussian corrected flash in AstroImageJ, using the star Gaia DR2 4094972869712553088 as the flux reference.

The coordinates of the lunar impact flash were obtained by performing geolocation through visual comparisons between the impact frames and LRO WAC images, using USGS lunar maps for reference.

\subsection{Crater Detection}
In total, PyNAPLE searched an area of 2$^o \times$ 2\degree centred on the estimated selenographic coordinates of latitude, $\phi$=8.0, longitude, $\lambda$=-76.5. 169 images were evaluated for suitability, 100 "before" and 69 "after" images, from which 82 temporal pairs were formed covering ~48\% of the search space.

\begin{figure*}
	\includegraphics[width=\textwidth]{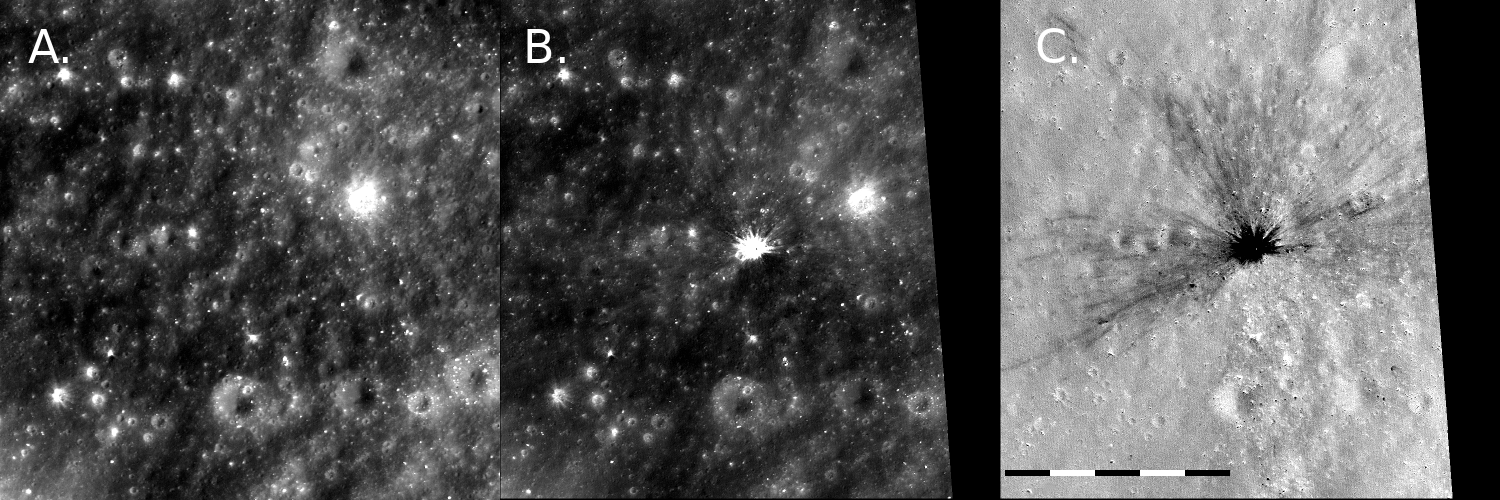}
	\caption{\textbf{(A)} The "before" image, M1180620010R, taken on 2015-03-10. \textbf{(B)} The "after" image, M1315871095R, taken on 2019-06-22, \textbf{(C)} The ratio of the two images. All three images are at the same scale of 1.12~m~pix$^{-1}$, the scale bar is 500~m, and each segment 100~m. Cropped from PyNAPLE processed NAC images from NASA/GSFC/Arizona State University.}
	\label{fig:crater_img}
\end{figure*}

After examining the resultant ratio images, a candidate crater was discovered at latitude, $\phi$=8.0288, longitude, $\lambda$=-76.546 in three images, M1315871095R, M1344064055L, and M1359341218L; Fig.~\ref{fig:crater_img} shows the crater imaged in the temporal pair of M1180620010R and M1315871095R. 

While there is an interval of 1156 days between the latest "before" image and the earliest "after" image, due to the fact the crater is observed in all images taken later than 2017-09-27, and is not present in any images from before 2017-09-27, it is a reasonable assumption that the crater was resultant from the observed 2017-09-27 LIF.

Further evidence supporting this hypothesis is that an impact flash of this duration occurs much less frequently than typical 1-2 frame (a few 10s of milliseconds) LIFs more commonly observed \citep{suggs2014,avdellidou2021}. As the impact crater has a much greater extent than typically observed newly formed impact craters, the coincidence of these being independent events has a low probability. 

\section{Discussion}

Due to the nature of the LRO's orbit, the time between subsequent images of a point on the lunar surface can be multiple years. This means that while the LROC collects a high volume of image data, the available images for a given point can be scarce. This scarcity can be exacerbated further when forming "before" and "after" image lists when the event is only a few years old.
Consequently, in the case of our identified crater, the number of "after" images of the formed crater is limited, and in each of the available "after" images the crater rim is unresolvable above the high albedo of the freshly excavated regolith. This limits the analyses that can be performed for the refinement of crater scaling laws using this impact event, until more images are obtained of the impact site where the crater morphology and rim diameter can be observed.

As an upper limit for crater size can be obtained, however, and morphology can be inferred from the ejecta pattern. The main contiguous ejecta blanket is elliptical with a major axis diameter of 90~m, and a minor axis diameter of 65~m. From these measurements, and the relationship for continuous ejecta size from crater diameter by \cite{moore1974}, shown in Eq.~\ref{eq:ejecta}, we can estimate the rim-to-rim diameter of the crater to be between 27.3-37.7~m.

\begin{equation}
    R_{ce} = 2.348R^{1.006}
	\label{eq:ejecta}
\end{equation}
where $R_{ce}$ is the radius of the continuous ejecta blanket, and $R$ is the radius of the crater. 

The asymmetrical nature of the ejecta blanket indicates that the impact was oblique; the butterfly ejecta pattern with an up-range exclusion zone indicates that the angle of impact $\theta_i < 30^o$, and the lack of a down-range exclusion zone indicates $\theta_i > 10^o$ \citep{melosh1989}.

In order to achieve the desired link between impactor and crater, the physical properties of the former upon the impact should be estimated. In recent years there have been methods developed to estimate the impact speed, $V$, as well as the mass of the impactor, $m$ \citep{suggs2014,madiedo2015b,avdellidou2021}. The luminous energy, $E_{lum}$, radiated by the flash and recorded at each frame during the observations is calculated as:

\begin{equation}
    E_{lum} = 3.75\times10^{-8}\times10^{-\frac{mag}{2.5}}\pi f\Delta\lambda D^2 t
	\label{eq:power}
\end{equation}
where $mag$ is the apparent magnitude at each frame, $f$ is a unit-less parameter denoting the isotropy of the flash, taken here as $f$=2, denoting the light originated from the lunar surface, $\Delta\lambda$ is the wavelength range, $\Delta\lambda$~=~300~nm, and $D$ is the Earth-Moon distance. Integrating the $E_{lum}$ for all the time that the flash was observed we estimated that the total $E_{lum}$ released was 1.3$\times10^7$~J. As mentioned before, the released E$_{lum}$ is a small fraction $\eta$ (luminous efficiency) of the total $KE$ of the impactor and is given by:
\begin{equation}
	KE = \frac{E_{lum}}{\eta}
	\label{eq:lum}
\end{equation}

Using two frequently-used estimations for the luminous efficiency $\eta_{1}$=5$\times$10$^{-4}$ and $\eta_{2}$=1.5$\times$10$^{-3}$ \citep{bouley2012}, we calculated the $KE$ of the impacting meteoroid to be $KE_1$=2.6$\times$10$^{10}$~J, and $KE_2$=8.7$\times$10$^{9}$~J respectively.

Knowing the $KE$, the next step is to estimate the impact speed $V$. By computing the angle between known active meteoroid streams at the time of impact and the location of the impact, an estimate can be made as to the parent stream responsible for the impactor. While it cannot solve for certain the parent stream of the meteoroid, some exclusions can be made to eliminate candidate streams. 
Meteoroid streams not active at the time of impact can be immediately eliminated as candidates. Meteoroid streams which have a radiant more than 90\degree offset from the normal of the impact can be eliminated, as the gravitational effect of the Moon on the trajectory is negligible, and therefore only streams with a radiant visible to the impact site are valid. If after exclusions there are no remaining candidate parent streams, the impactor belongs to the sporadic meteoroid background.

The active streams during the 2017-09-27 impact are shown in Tab.~\ref{tab:streams}. Due to the radiant being >90\degree from the impact site, the Daytime Sextantids can be immediately excluded. Both the Southern Taurids, and Southern Delta Piscids have the radiant visible to the impact site, and can both be considered candidate streams. Usually, in order to distinguish between two candidate streams, the stream with the smaller angular separation between the lunar solar longitude, and the solar longitude of the meteoroid streams peak. This parameter can be somewhat ignored if one of the streams has a much higher Zenithal Hourly Rate (ZHR) and the separation between stream peaks is not large. 

There is some debate over whether the Southern Taurids and Southern Delta Piscids are distinct streams, with indication they could in-fact originate from the same source \citep{triglav-cekada2005}. Consequently impactors from these two streams would comprise of similar materials with similar densities. Due to the streams having $\Delta$$V_G$=2.0~km~s$^{-1}$, and the radiant of the two streams having an angular separation of only 11\degree, distinguishing between these streams would have a negligible effect on any subsequent calculations. Thus we used the average speed of the streams $V_G$=27.6~km~s$^{-1}$ for the mass estimation.

However, because the parent stream was not determined for certain and we cannot exclude the possibility the impactor to originate from the background population of the meteoroids, we estimate its mass also in this case. Several average values for the speed of the background population have been adopted so far from the different LIF studies, ranging from 16 to 24~km~s$^{-1}$ \citep{steel1996,ivanov2006,mcnamara2004,lefeuvre2011}. Here we adopt an intermediate one of $V$=20.0~km~s$^{-1}$.

Using Eq.~\ref{eq:lum} we estimated the mass of the impactor to be between 22.8 and 130~kg, following both scenarios where the impactor originates from a stream or from the background population, repeated for the two different $\eta$ values. 

\begin{table}
	\centering
	\caption{The active streams at the time of the 2017-09-27 impact. $V_G$ is the group velocity of the stream. Values for solar longitude, $\lambda_{\odot}$, and $V_G$ are taken from \protect\cite{jenniskens2016}.}
	\label{tab:streams}
	\begin{tabular}{lccccr} 
		\hline
		Stream					& Code 	& $\lambda_{\odot}$ & ZHR 	& $V_G$ (km~s$^{-1}$)\\
		\hline
		Daytime Sextantids		& DSX 	& 188\degree 			& 20	& 32.9 \\
		Southern Taurids		& STA 	& 216\degree 			& 5		& 26.6 \\
		Southern Delta Piscids	& SPI 	& 176\degree 			& 3		& 28.6 \\
		\hline
	\end{tabular}
\end{table}




The next step was to use the impactors parameters and utilising an impact scaling law to predict the diameter of the crater. The crater scaling law by Melosh, described in Eqs.~\ref{eq:Dscale} and \ref{eq:Gscale}, was selected as it is the most recent and is frequently-used in recent lunar impact analysis \citep{schmidt1987,melosh1989}.

\begin{equation}
    D_c = \gamma^{-0.26}m^{0.26}V^{0.44}
	\label{eq:Dscale}
\end{equation}
\begin{equation}
    \gamma = 0.31g^{0.84}\rho_p^{-0.26}\rho_t^{1.26}{\left(\frac{sin 45}{sin \theta}\right)}^{1.67}
	\label{eq:Gscale}
\end{equation}
where $D_c$ is the crater diameter, $m$ is the impactor mass, $V$ is the impactor velocity, $g$ is the gravitational acceleration, $\rho_p$ is the impactor density, $\rho_t$ is the target density, $\theta$ is the impact angle with respect to the vertical. The units are in the MKS system.

Following the same logic, $D_c$ calculations were performed for both cases of impactors origin; the sporadic background population, and for the STA/SPI meteoroid streams. As previously discussed, the STA and SPI meteoroid streams are similar in velocity, direction and projectile density. For this reason, a nominal value of $V$=27.6~km~s$^{-1}$ was used. The projectile density was taken as $\rho_p$=2,500~kg~m$^{-3}$, the bulk density of the STA meteoroids \citep{matlovic2017}. For calculations assuming a sporadic background meteoroid, $V$=20.0~km~s$^{-1}$ and $\rho_p$=1,800~kg~m$^{-3}$ were used. 
For both calculations target the bulk density of lunar soil was used, $\rho_t$=1,500~kg~m$^{-3}$, and the value of lunar gravity $g$=1.625~m~s$^{-2}$ was used. A minimum and maximum impacting angle were taken as $\theta_i$=10\degree and $\theta_i$=30\degree respectively, as denoted by the range of angles possible for the morphology of the observed crater.

Assuming an STA/SPI and sporadic origin of the meteoroid, calculations estimated a $D_c$ between 39.4-55.3~m and 24.1-50.7~m respectively, Tab.~\ref{tab:diams} contains a more in depth breakdown of this range. Both of these ranges are in agreement with the crater diameter estimated from the continuous ejecta blanket (27.3-37.7~m), showing that the small difference in the impact speed did not affect the result. However, further assessment to distinguish the most likely candidate would require unsaturated images of the crater. 

As PyNAPLE has now located its first linked crater from an observed impact flash, a larger scale implementation of the PyNAPLE algorithm can commence. By utilising the publicly-available lunar impact flash data collected since 2009 (LROs mission start), a wide scale mission to locate and identify the linked craters can be performed.
This is done with the intention of forming a dataset of craters for which the impact energy and resultant size are precisely known, therefore enabling a statistical analysis of the craters formed. Furthermore, discovering and measuring craters for which we also have the physical properties of their impactor will allow the more precise estimation of the luminous efficiency, which currently may vary by an order of magnitude. A fresh crater could also act as a good surface exploration site to study how space weathering affects fresh craters in the short term.

\begin{table}
	\centering
	\caption{The diameter of the resultant impact crater, calculated using the crater scaling law described in Eq.~\ref{eq:Dscale} \& \ref{eq:Gscale} for each luminous efficiency $\eta_1=5\times10^{-4}$ and $\eta_2=1.5\times10^{-3}$ using the maximum and minimum impacting angles for the observed crater morphology.}
	\label{tab:diams}
	\begin{tabular}{lc|cc|cr}
		\hline
		Origin & $\theta_{i}$ & D$_{\eta1}$ & D$_{\eta2}$ \\
		\hline
		SPI/STA	& 10\degree & 55.3~m & 41.6~m \\
				& 30\degree & 52.3~m & 39.4~m \\
		Sporadic& 10\degree & 32.0~m & 24.1~m \\
				& 30\degree & 50.7~m & 38.1~m \\
				
		\hline
	\end{tabular}
\end{table}

\section{Conclusions}

We presented a fully automated pipeline for the collation, formation, and processing of LRO NAC temporal pairs. PyNAPLE is the first algorithm designed to detect fresh craters using the coordinates and epoch of lunar impact flashes. PyNAPLE is fully configurable, offering the ability to fine tune the matching algorithms, map projections, and tolerances of the data selection and pattern matching stages to optimize PyNAPLE for individual needs. 
We discussed the trio of tests performed on PyNAPLE, and the successful location of a newly formed crater with high probability of being formed from the flash.

Preliminary analyses were performed on the 2017-09-27 linked impact crater, allowing us to conclude that the impact occurred at an angle of between 10-30\degree, forming a crater predicted by scaling laws to be between 24.1-55.3~m in size, and the ejecta blanket to crater rim relationship predicting a diameter between 27.3-37.7~m.
This finding demonstrated that PyNAPLE is a functioning tool for the detection of new craters from new lunar impact flash data. 
\section*{Acknowledgements}

This work was supported by the Programme National de Planetologie (PNP) of CNRS/INSU, co-funded by CNES and by the program "Flash!" supported by Credits Scientifiques Incitatifs (CSI) of the Universite Nice Sophia Antipolis. CA was supported by the French National Research Agency under the project "Investissements d' Avenir" UCA$^{JEDI}$ with the reference number ANR-15-IDEX-01. DJS, CA and MD acknowledge support from the \textit{ANR "ORIGINS"} (ANR-18-CE31-0014).

This work presents results from the European Space Agency (ESA) space mission Gaia. Gaia data are being processed by the Gaia Data Processing and Analysis Consortium (DPAC). Funding for the DPAC is provided by national institutions, in particular the institutions participating in the Gaia MultiLateral Agreement (MLA). The Gaia mission website is https://www.cosmos.esa.int/gaia. The Gaia archive website is https://archives.esac.esa.int/gaia.

ISIS3 software courtesy of the U.S. Geological Survey.

%

\section*{Data Availability}
With PyNAPLE's completion it is now publicly available at \url{github.com/dsheward-astro/PyNAPLE} for both professionals and amateurs to use.
%



\bibliographystyle{mnras}
\bibliography{references.bib} 







\bsp	
\label{lastpage}
\end{document}